\documentstyle[12pt]{article}
\input math_macros.tex

\begin{document}
\begin{titlepage}
\begin{center}

May 15, 1998     \hfill    LBNL-41813 \\

\vskip .5in

{\large \bf Is Quantum Mechanics Non-Local?}
\footnote{This work was supported by the Director, Office of Energy 
Research, Office of High Energy and Nuclear Physics, Division of High 
Energy Physics of the U.S. Department of Energy under Contract 
DE-AC03-76SF00098.}
\vskip .50in
Henry P. Stapp\\
{\em Lawrence Berkeley National Laboratory\\
      University of California\\
    Berkeley, California 94720}
\end{center}

\vskip .5in

\begin{abstract}

A recent proof, couched in the symbolic language of modal logic, shows 
that a well-defined formulation of the question posed in the 
title is answered affirmatively. In a paper with the same title Unruh has 
tried in various ways to translate the symbolic proof into normal prose,
and has claimed that the proof must fail in some way or another. A correct 
translation is given here, and it is explained why the difficulties 
encountered by Unruh do not actually arise.

\end{abstract}
\medskip
\end{titlepage}

\renewcommand{\thepage}{\roman{page}}
\setcounter{page}{2}
\mbox{ }

\vskip 1in

\begin{center}
{\bf Disclaimer}
\end{center}

\vskip .2in

\begin{scriptsize}
\begin{quotation}

This document was prepared as an account of work sponsored by the United
States Government. While this document is believed to contain correct 
 information, neither the United States Government nor any agency
thereof, nor The Regents of the University of California, nor any of their
employees, makes any warranty, express or implied, or assumes any legal
liability or responsibility for the accuracy, completeness, or usefulness
of any information, apparatus, product, or process disclosed, or represents
that its use would not infringe privately owned rights.  Reference herein
to any specific commercial products process, or service by its trade name,
trademark, manufacturer, or otherwise, does not necessarily constitute or
imply its endorsement, recommendation, or favoring by the United States
Government or any agency thereof, or The Regents of the University of
California.  The views and opinions of authors expressed herein do not
necessarily state or reflect those of the United States Government or any
agency thereof or The Regents of the University of California and shall
not be used for advertising or product endorsement purposes.
\end{quotation}
\end{scriptsize}

\vskip 2in

\begin{center}
\begin{small}
{\it Lawrence Berkeley Laboratory is an equal opportunity employer.}
\end{small}
\end{center}

\newpage
\renewcommand{\thepage}{\arabic{page}}
\setcounter{page}{1}

One of the great lessons of quantum theory is that utmost caution must be 
exercised in reasoning about hypothetical outcomes of unperformed 
experiments. Yet Bohr [1] did not challenge the argument of Einstein,
Podolsky, and Rosen [2] on the grounds that it was based on the 
simultaneous consideration of mutually exclusive possibilities.
Rather he challenged the underlying EPR presumption that an 
experiment performed locally on one system would occur ``without in 
any way disturbing '' a faraway system. Bohr's own ideas rested 
heavily on the idea that experimenters could freely choose between
alternative possible measurements, and the core of his answer to EPR was that 
although {\it ``... there is in a case like that just considered no 
question of a mechanical influence of the system under investigation during 
the last critical stage of the measuring procedure.''}...``there is 
essentially the question of {\it an influence of the very conditions which 
define the possible types of predictions about future behavior of the 
system.''} 

The adequacy of Bohr's answer, and the nature of his intermediate 
position on the question of these influences, has been much debated.
The issue is of fundamental importance, because it concerns the nature 
of the causal structure of quantum theory, and its compatibility with the 
idea, drawn from the theory of relativity, that no influence of any kind can 
act backward in time in any frame.

To obtain rigorous results in this realm it is useful to formulate
arguments within a formal logic, where
each separate statement can be stated precisely, and the rules of inference
connecting them are exactly spelled out. This was done in reference [3] 
where a logical contradiction was shown to arise from a combination of 
explicitlty stated assumptions. One of these assumptions was a locality 
(no-faster-than-light-influence) condition LOC1, which expresses
Bohr's ``no mechanical influence'' condition. It states that if an
experiment L2 is actually performed in a spacetime region L,
and an experiment R2 is actually performed in a faraway region R that 
lies later in time than L (in some frame), and if an outcome c appears to the 
observers stationed in the earlier region L, then that same result c would 
have appeared to the observers in L in the alternative possible case in 
which nothing is changed except for the later-in-time free choice by the 
experimenter in R and the consequences of this change: LOC1 asserts that
the later free choice in R has no effect on the outcome that appeared 
earlier to the observers stationed in region L.

The argument in reference [3], stated here in words, rather than the 
symbols of modal logic, starts as follows:

Suppose the actual situation is such that L2 and R2 are performed and the
outcome g appears to the observers in R. Then a prediction of quantum 
theory, in the Hardy case under consideration in [3], entails that the 
outcome appearing to the observers in L must be c, in this actual situation. 
But according to LOC1 this actual outcome c appearing to the observers in L 
would not be disturbed if the later free choice in R were to perform R1 
instead of R2: the actual outcome appearing to the observers in L would remain
c. But then if the laws of quantum theory are assumed to hold not only in 
the single unique world that is actually created by our free choices but also 
in the alternative possible worlds that would be created if our alternative 
free choices had been different, then another prediction of quantum theory 
in this Hardy case entails that the outcome appearing to the observers in R, 
in this alternative possible situation (in which R1 is performed instead of 
R2) must be f. 

This result is summarized in line 5 of my proof, which in prose reads:

LINE 5: If L2 is performed then SR is true,

\noindent where SR is the statement:

SR: ``If R2 is performed in R and outcome g appears to the observers in R, then
if R1, instead of R2, had been performed in R the outcome f would have appeared
to the observers in R.''

The form of this claim in line 5 is the same as a typical claim in classical
mechanics: if the result of a certain measurement 1 is, say, g, then the 
deterministic laws of physics may allow one to deduce that if some alternative
possible measurement 2 had been performed, instead of 1, then the result of 
that measurement 2 would necessarily have been f: knowledge of what happens 
in an actual experimental situation may, with the help of theory, allow one 
to infer what would have happened if one had performed, instead, a different
experiment.

Note that the assumptions in line 5 pertaining to region L do not include 
the condition that outcome c appears to the observers in L: that condition 
is implied by the predictions of quantum theory and the explicitly 
stated conditions, namely that L2 and R2 are actually performed, and that 
outcome g appears to the observers in R.

This prose description is clear and direct, and it conforms to the meanings 
ascribed to the statements in my proof by the rules modal logic. This logic 
has been developed by philosophers and logicians for dealing with statements 
of the sort occurring here. Of course, that logic was designed to formalize
normal rational thinking, and it may therefore incorporate philosophical 
prejudices drawn from the prevailing classical conception of the nature of 
reality. Still,  Bohr strongly supported the idea that no non-classical-type of
logic or reasoning is needed to deal with our descriptions at the level of 
possible experimental set-ups, and our observations of their outcomes. No 
other kind of description enters into my argument. 

Unruh [4] has set forth an array of different interpretations of 
practically every statement in my proof, and has encountered all sorts 
of difficulties. 
	
Of the interpretations of LOC1 offered by Unruh, the one closest to the one 
occurring in my proof is the one he describes first: ``On face value this is 
just the unexceptional statement that if L2 is measured to have value c then 
the truth of having obtained that value is independent of what is (or 
will be) measured at R''. My statement is only slightly different: 
``This is just the unexceptional statement that if L2 is measured and the 
outcome appearing to the observers in L is c, then this latter fact is 
independent of which measurement will later be performed in R: the later free 
choice by the experimenter in R does not disturb what was observed earlier 
by the observers in region L.'' 

Unruh claims that ``this meaning of LOC1 is insufficient to derive his 
[Stapp's] conclusion, since it demands that L2 had actually been measured 
and had the given outcome. It ties the meaning of this locality condition to 
the actuality of the measurement and its outcome. This does not allow 
counterfactual replacement of L2, since the truth of this statement is, 
under this interpretation, tied to the truth of the measurement 
of attribute L2 actually having been carried out in L, and having obtained 
that specified outcome.'' [I have replaced by L2 some apparent misprints]

To evaluate this objection we must turn to the second locality condition,
LOC2, and its application, for that is where the condition L2 is relaxed.

LOC2 is connected to line 5, which says that if L2 is performed then statement 
SR is true.

Suppose the experiment L2 is performed in a spacetime region that lies much
later in time (in some frame) than all points in the region R in which all 
of the possible events referred to by SR lie. And suppose the later choice 
between L1 and L2 is really free: i.e., that this choice is independent of 
everything that happens earlier. 

Then the demand that there be no backward-in-time influence of any kind
requires that SR cannot be necessarily true if the later free choice
is L2, but be untrue if the later free choice is L1: such a difference would
constitute {\it some sort} of backward-in-time influence.

LOC2 is, accordingly, the assertion that if SR is true under condition L2,
then it is also true if L1, instead of L2, is performed in L.

Of course, the {\it proof} that SR is true under condition L2 depends upon
the fact that L2 is performed, and that {\it proof} certainly fails if
L2 is not performed. But LOC2 does not say that the {\it proof} of SR
carries over to the case in which L1, instead of L2, is performed in L. 
It asserts only the {\it truth} of SR does not depend on a free choice that 
is made at a time that is later than the times of all the possible events 
whose occurrences or non-occurrences under various conditions specified in R 
determine, by definition, the truth of SR.

What does Unruh say about LOC2?

He says: ``If it were true that one could deduce solely from the fact that 
a measurement had been made at L that some relation on the right must hold,
then I would agree that this requirement [LOC2] would be reasonable.''

Well, one needs, of course, the other assumptions, including LOC1. But, given 
those other assumptions, which I have specified, one can deduce SR ``solely'' 
from the assumption that L2 is performed: one does not need to assume that
the outcome appearing to the observers stationed in L is c. So 
Unruh's statement appears to confirm that LOC2 is a valid expression 
of the condition that there be no influence backward in time.

But then he immediately says: ``However, if the truth of the relation on 
the right hand side depended not only on which measurement had been made 
[I would say ``will be made''] on the left, but also on the actual value 
obtained on the left, then no such locality condition would obtain.'' 

This muddies the waters, for it seems to contradict what he just said.
How is this conflict resolved?

In logic one distinguishes between ``proof'', ``truth'', and ``meaning''.
Unruh's arguments blur these distinctions. He speaks of the dependence
of the {\it truth} of SR upon actual values obtained on the left. 
What is true is that the {\it proof} of SR depends on the actual value c
that appears on the left. But it does not follow that the {\it truth}
of SR depends on what happens later in region L. The truth of SR depends,
by definition, on the occurrence or non-occurrence of possible events in the 
earlier region R, under various conditions freely choosen in R. LOC2 is the
assertion that the free choice made later by the experimenter in L leaves
undisturbed an existing pattern in those earlier possible events in R.  
 
What about Unruh's earlier claim that ``this meaning of LOC1, is 
insufficient to derive his [Stapp's] conclusion, since it demands that L2 had 
actually been measured and had the given outcome. It ties the meaning of this 
locality condition to the actuality of the measurement and its outcome. This 
does not allow counterfactual replacement of L2, since the truth of this 
statement is, under this interpretation, tied to the truth of the measurement 
of attribute L2 actually having been carried out in L, and having obtained 
that specified outcome.''

It is of course essential, in following a logical argument, to proceed 
step-by-step in a logical progression that leads from the assumptions to the 
conclusions, rather than collapsing the steps to a single unjustified leap.
In my proof the assumption LOC1 is used to get line 5. In that application
of LOC1 the experiment performed in region L is fixed to be the actually 
performed experiment L2. And the premise of SR, together with a 
prediction of quantum theory, allows one to conclude that the outcome 
actually observed by the observers in L is the outcome c. So all the 
conditions for the applicability of LOC1 are satisfied. LOC1 is not used 
thereafter. 

The answer, therefore, to Unruh's claim about the insufficiency of LOC1
is that LOC1 is used only under conditions where its use is justified, 
and in particular only under the condition that L2 is performed in L. 

Unruh advances several other arguments that I should also address.
He suggests that perhaps LOC1 means that ``if one is somehow able to
infer that L2 has value c then it remains true that L2 has value c
under replacement of R2 with R1 even if the outcome of the measurement
R2 was crucial in drawing the inference that L2 has value c.'' He says that:
``This interpretation of LOC1 is. I would argue, a form of realism, in that it 
claims that the value to be ascribed to L2 is independent of the evidence 
used to determine that value.''

This argument confounds several separate steps in my argument. LOC1 asserts 
merely that if L2 and R2 are performed in L and R respectively, where L is 
earlier than R in some frame, and if outcome c appears to the observers 
stationed in L, then that same result c would appear to these observers in L 
if everything were kept exactly the same except for the later-in-time free 
choice by the experimenter in R, and the consequences of that change.  In the 
actual situation specified by the three conditions L2 and R2 and g a result 
from quantum theory entails that the outcome c appears to the observers in L. 
Given this fact, one can use LOC1 to deduce what ``would happen'' in L in 
the hypothetical situation in which this factual situation is changed solely 
by a change in the free choice in R and the consequences of that change. 
According to LOC1 this change in R would not disturb the actual outcome that 
appears to the observers stationed in the earlier region L: the observers 
in the earlier region L would continue to observe outcome c.

As in most logical arguments it is important to follow the steps in the proof. 
In this case the ``evidence'' for the fact that the actual result in L is c, 
lies in the fact that the actual situation under consideration is one in 
which L2 and R2 are performed, and the outcome appearing to the observers in R
is g, combined with the knowledge provided by quantum theory that under these 
conditions the outcome appearing to the observers in L is c. 

Of course, one of the consequences of changing the free choice made in
R from R2 to R1 is that under these altered conditions the outcome 
appearing to the observers {\it in R} would be different from what it 
actually is. But LOC1 asserts that what has already been observed by the 
observers stationed in the earlier region L will not be changed by this 
change in the later free choice made in R: the present actual evidence of 
one's senses does not depend on what some experimenter will freely choose 
to do tomorrow, as it might well do if that later free choice could effect 
prior events.
 
Unruh considers my line 4, and says: ``The claim is that this arises
solely by logic out of the substitution of (2) and (3) into (1). However, it 
is now clear that the meanings ascribed to LOC1 and to the statements of the 
logical calculus are crucial.'' 

It certainly is clear that meanings ascribed to LOC1 and the statements of the
logical calculus are crucial. I have spelled out here in prose what these 
meanings are, and how they enter into my proof. Unruh proposes some other 
meaning, which he formalizes in his Eq. 14. 

Whatever that equation means, it definitely does not
conform to the rules of modal logic. Certainly, one  cannot say, as Eq. 14
does, that if R1 is performed instead of R2 then R2 is performed. And Unruh's
claim that the final inference, namely that (L2 and c and R2 and f) implies 
(L2 and R1 and f) ``is no longer true'' certainly makes his logic 
incompatible with modal logic, and with classical logic as well.

Unruh goes on to restate his objections in various other ways,
But I believe I have already explained how my proof rationally 
evades all of the difficulties he encountered.

\end{document}